\magnification \magstep1
\raggedbottom
\openup 2\jot
\voffset6truemm
\def\pd{\partial}
\def\su{\over}
\def\a{\alpha}
\def\px{\phantom{i}} 
\def\pdx{\pd\px}
\def\th{\theta}
\headline={\ifnum\pageno=1\hfill\else
{MULTIMOMENTUM MAPS ON NULL HYPERSURFACES}
\hfill \fi}
\centerline {\bf MULTIMOMENTUM MAPS ON}
\centerline {\bf NULL HYPERSURFACES}
\vskip 1cm
\noindent
{GIAMPIERO ESPOSITO and COSIMO STORNAIOLO}
\vskip 1cm
\noindent
{\it Istituto Nazionale di Fisica Nucleare, Sezione di Napoli,
Mostra d'Oltremare Padiglione 20, 80125 Napoli, Italy}
\vskip 0.3cm
\noindent
{\it Dipartimento di Scienze Fisiche, Mostra d'Oltremare
Padiglione 19, 80125 Napoli, Italy}
\vskip 1cm
\noindent
{\bf Summary.} -- This paper studies the 
application of multimomentum maps to
the constraint analysis of general 
relativity on null hypersurfaces. It is
shown that, unlike the case of spacelike hypersurfaces,
some constraints which are second class in the
Hamiltonian formalism turn out to contribute
to the multimomentum map.
To recover the whole set of secondary constraints found in the 
Hamiltonian formalism, it is necessary to combine the
multimomentum map with those particular Euler-Lagrange equations 
which are not of evolutionary type. The analysis is performed
on the outgoing null cone only.
\vskip 6cm
\leftline {PACS numbers: 04.20.Cv, 04.20.Fy}
\vskip 100cm
\leftline {\bf 1. - Introduction.}
\vskip 0.3cm
\noindent
In the Hamiltonian formulation of general relativity, the
constraint analysis on null hypersurfaces 
plays an important role
since such surfaces provide a natural framework for the study
of gravitational radiation in asymptotically flat space-times [1-6].
Moreover, in a null canonical formalism, the physical
degrees of freedom and the observables of the theory may be
picked out more easily [4,5]. 

On the other hand, relying on the multisymplectic formalism
for classical field theories described, for example, in ref.[7],
recent work in the literature [8-10] has studied the formulation
of general relativity in terms of jet bundles. In this 
formalism, the local description involves local coordinates
on Lorentzian space-time, tetrads, connection one-forms,
multivelocities corresponding to the tetrads and multivelocities
corresponding to the connection one-forms. The derivatives of
the Lagrangian with respect to the latter class of 
multivelocities give rise to a set of {\it multimomenta}
which naturally occur in the constraint equations. All the
constraint equations of general relativity are then found to
be linear in terms of this class of multimomenta. In ref.[9], the
construction of ref.[8] has been 
extended to complex general relativity,
where Lorentzian space-time is replaced by a four-complex-dimensional
complex-Riemannian manifold. One then finds a holomorphic theory
where the familiar constraint equations are replaced by a set
of equations linear in the holomorphic multimomenta, provided
that such multimomenta vanish on a family of two-complex-dimensional
surfaces [9,10].

In the light of the properties and results briefly outlined,
we have been led to consider the Lagrangian version
of a constraint analysis on null hypersurfaces, when the
multisymplectic formalism [7] is applied. For this purpose,
sect. {\bf 2} describes null tetrads, 
while the analysis of multimomentum maps on null hypersurfaces
is performed in sect. {\bf 3}. Self-dual gravity is studied in
sect. {\bf 4}, and 
concluding remarks are presented in sect. {\bf 5}.
\vskip 10cm
\leftline {\bf 2. - Null tetrads.}
\vskip 0.3cm
\noindent
In this paper we are only interested in a local analysis of
null hypersurfaces. Thus,
many problems arising from the possible null-cone
singularities are left aside. To give a geometric 
description of a null hypersurface, it is possible to introduce,
as in ref.[4], a null tetrad with components
$$
e_{\hat 0}={1\over N}\left({\pdx\over\pd t} - N^{i}{\pdx\su\pd 
x^{i}}\right) \; ,
\eqno (2.1)
$$
and
$$
e_{\hat k}=
-{\a_{\hat k}\over N} {\pdx\su\pd t} 
+\left(V_{\hat k}^{i} 
+ \a_{\hat k} {N^{i}\over N} \right){\pdx\su\pd x^{i}} \; ,
\eqno (2.2)
$$
where $N$ is the lapse function and $N^{i}$ are components
of the shift vector. The duals to (2.1) and (2.2) are
$$
\th^{\hat 0}=(N+\a_{i} N^{i})dt + \a_{i} dx^{i} \; ,
\eqno (2.3)
$$
and
$$
\th^{\hat k}=\nu^{\hat k}_{i} \Bigr(N^{i}dt+dx^{i}\Bigr) \; , 
\eqno (2.4)
$$
where tetrad labels $\hat a$, $\hat b$, $\hat c=0,1,2,3$,
while the indices $\hat k$, $\hat l=1,2,3$. 
Analogous notation is used for the space-time indices $a$,$b$,... 
and $i$, $j$. Moreover, one has
$$
V_{\hat k}^{i} \; \nu^{\hat l}_{i}=\delta^{\hat l}_{\hat k} \; ,
\eqno (2.5)
$$
and
$$
\a_{\hat k}=V^{i}_{\hat k}\a_{i} \; .
\eqno (2.6)
$$
Given the metric defined by
$$
\eta_{\hat a\hat b}=\eta^{\hat a\hat b}
\equiv \left(\matrix{0&1&0&0\cr
1&0&0&0\cr 
0&0&0&-1\cr
0&0&-1&0\cr}\right) \; ,
\eqno (2.7)
$$
the space-time metric can be expressed as
$g=\eta_{{\hat a}{\hat b}}\theta^{\hat a} \otimes \theta^{\hat b}$.
It is then straightforward to see that, on the hypersurfaces
defined locally by the equation $t={\rm constant}$, one has 
$$
g^{ab}t_{,a}t_{,b}=-{2\su
N^2}\,(\a_{\hat 1} + \a_{\hat 2} \a_{\hat 3}) \; .
\eqno (2.8) 
$$ 
This implies that
such hypersurfaces are null if and only if 
$$
\a_{\hat 1} +
\a_{\hat 2} \a_{\hat 3}=0 \; .
\eqno (2.9)
$$ 
By a particular choice of coordinates, it is
always possible to set $\a_{\hat 2}= \a_{\hat 3}=0$ [4].

In most of the following equations the tetrad vectors appear in the 
combination
$$
{\widetilde p}^{\; \,ac}_{\; \phantom{\,ac} \hat a\hat c } 
= {e\over 2} \left(e^a_{\hat a}
e^{c}_{\hat c} - e^{a}_{\hat c}e^{c}_{\hat a} \right) \; ,
\eqno (2.10)
$$
where $e=N\nu$ with $\nu={\rm det}(\nu^{\hat a}_i)$. In a covariant 
Hamiltonian version of the theory, these quantities can be identified with 
the multimomenta introduced in refs.[8,10].
\vskip 0.3cm
\leftline {\bf 3. - Multimomentum maps on null hypersurfaces.} 
\vskip 0.3cm
\noindent
The multimomentum map is a 
geometric tool which encodes the relevant information about
the invariance properties of a classical field theory and
its first-class constraints [7--10]. Indeed, the terminology
used so far by the authors [8--10] differs from the one in 
ref.[7]. As far as we can see, what we call multimomentum
map corresponds to the {\it energy-momentum map} 
defined in ref.[7].

In particular, in general
relativity, the evaluation of the multimomentum map on
a spacelike hypersurface $\Sigma$ can be expressed in terms
of the following integral [8]:
$$ 
I_{\Sigma}[\xi,\lambda]=\int_{\Sigma}\left[
{\widetilde p}_{\; \; \; \; {\hat b}{\hat d}}^{\; ac}
\biggr(\xi_{\; \; ,a}^{b}
\; \omega_{b}^{\; \; {\hat b}{\hat d}}
-(D_{a}\lambda)^{{\hat b}{\hat d}}
+\omega_{a \; \; \; \; ,b}^{\; \; {\hat b}{\hat d}}
\; \xi^{b}\biggr) 
+{1\over 2} {\widetilde p}_{\; \; \; \; {\hat b}{\hat d}}^{\; ab}
\; \Omega_{ab}^{\; \; \; {\hat b}{\hat d}}
\; \xi^{c}\right]d^{3}x_{c} \; .
\eqno (3.1)
$$
With our notation, $\xi$ is a vector field 
describing infinitesimal diffeomorphisms on
the base space ({\it i.e.}, space-time), 
$\omega_{a}^{\; \; {\hat b}{\hat c}}$ are
the connection one-forms and $\Omega_{ab}^{\; \; \; {\hat c}{\hat d}}$ are
the curvature two-forms. Moreover, the antisymmetric $\lambda^{{\hat
a}{\hat b}}$ is an element of the algebra $o(3,1)$, and $D_{a}$ denotes
covariant differentiation with respect to a Lorentz connection which
annihilates the Minkowskian metric of the internal space [8]. 

The key point of our analysis is now the evaluation of
the integral (3.1) on a null hypersurface ${\cal S}_{N}$, 
and then the
interpretation of the resulting contributions in terms of
a subset of the constraint equations. Indeed, by virtue of
the formalism described in sect. {\bf 2}, the multimomenta on
a null hypersurface read
$$
{\widetilde p}_{\; \; \; \; {\hat 0}{\hat l}}^{\; 0i}
={e\over 2N}V_{\; \; {\hat l}}^{i}
\; ,
\eqno (3.2)
$$
$$
{\widetilde p}_{\; \; \; \; {\hat 0}{\hat l}}^{\; ij}
=-{e\over N} N^{[i} \; V_{\; \; {\hat l}}^{j]}
\; ,
\eqno (3.3)
$$
$$
{\widetilde p}_{\; \; \; \; {\hat k}{\hat l}}^{\; 0i}
={e\over N}V_{[{\hat k}}^{i} \;
\alpha_{{\hat l}]}=0
\; ,
\eqno (3.4)
$$
$$
{\widetilde p}_{\; \; \; \; {\hat k}{\hat l}}^{\; ij}
=eV_{\; \; {\hat k}}^{[i} \; V_{\; \; {\hat l}}^{j]}
+{e\over N} V_{[{\hat k}}^{[i} \; 
\alpha_{{\hat l}]} \; N^{j]}
=eV_{\; \; {\hat k}}^{[i} \; V_{\; \; {\hat l}}^{j]}
\; ,
\eqno (3.5)
$$
where we have used the freedom to set to zero two of the $\alpha$
parameters, jointly with eq. (2.9). 
We now integrate by parts in eq. (3.1) and we
impose the boundary conditions of ref.[8], according to which the
multimomenta or the gauge parameters 
$\xi^{a}$ and $\lambda^{\hat a \hat b}$
should vanish at the boundary $\partial \Sigma$. Moreover, we restrict
ourselves to the adapted coordinates 
for null hypersurfaces (cf.[8]), which
implies that only the integration $d^{3}x_{0}$ survives in eq. (3.1). 
This means that the integration is only performed on the outgoing
null cone [1,2]. 

Thus, in the light of eqs. (3.2)--(3.5), on setting to zero
the multimomentum map on a null hypersurface (this is what one
does on spacelike hypersurfaces to obtain the constraints [7])
one finds the equations
$$
\int_{{\cal S}_N} \lambda^{{\hat 0}{\hat k}}
\left[\pd_{i}
\Bigr({e\over N}V_{\hat k}^{i}\Bigr)
+ {e\over N}\omega_{i \hat k}^{\phantom{i \hat k}\hat l}
\; V_{\hat l}^{i}
\right]d^{3}x_{0} =0 \; ,
\eqno (3.6)
$$
$$
\int_{{\cal S}_N} \lambda^{{\hat k}{\hat l}}
\Bigg[\pd_{i}
\biggr({e\over N}V_{[{\hat k}}^{i} \;
\alpha_{{\hat l}]}\biggr) 
+{e\over 2N}\Bigr[
\omega^{\phantom{ik}\hat 0}_{i\hat k}V^{i}_{\hat l}
-\omega_{i\hat l}^{\phantom{il}\hat 0}V^{i}_{\hat k}\Bigr]
+{e\over N}
\omega^{\phantom{ik}\hat s}_{i\hat k}
\; V^{i}_{[{\hat s}} \; \alpha_{{\hat l}]}
+{e\over N}\omega_{i\hat l}^{\phantom{i\hat l}\hat s} 
\; V^{i}_{[\hat k}
\; \alpha_{\hat s]} \Bigg] d^{3}x_{0}=0 \; ,
\eqno (3.7)
$$
$$
\int_{{\cal S}_N} \left[e V_{\hat k}^{[i} \; V_{\hat l}^{j]}
\; \Omega_{ij}^{\; \; \; {\hat k}{\hat l}}
-{2e\over N} N^{[i} \; V_{\hat k}^{j]}
\; \Omega_{ij}^{\; \; \; {\hat 0}{\hat k}}\right]
\xi^{0} \; d^{3}x_{0}=0 \; ,
\eqno (3.8)
$$
$$
\int_{{\cal S}_N} \left[ {e\over N} V_{\hat k}^{i} \; 
\Omega_{ij}^{\; \; \; {\hat 0}{\hat k}}\right] 
\; \xi^{j} \; d^{3}x_{0} =0 \; .
\eqno (3.9)
$$
However, eqs. (3.6)--(3.9) are only a subset of the full
set of constraints in the theory (see below).

Indeed, the Palatini action 
$$
S_P\equiv {1\over2}\int_M d^4x\,e\,e^a_{\hat a}\, e^b_{\hat b} \;
\Omega^{\phantom{ab}\hat a\hat b}_{ab} \; ,
\eqno (3.10)
$$
leads to the Euler-Lagrange equations [8]
$$
G^{\hat c}_h\equiv
e^{b}_{\;\hat b}\left[\Omega^{\phantom{bh}\hat b\hat c}_{bh}-
{1\over2}e^{d}_{\;\hat a}e^{\hat c}_{\; h}
\Omega^{\phantom{bd}\hat b\hat a}_{bd}\right]=0 \; ,
\eqno (3.11)
$$
and
$$
D_{b} \; {\widetilde p}^{\;\, ab}_{\phantom{\;\,ab} 
\hat a \hat b}=0 \; .
\eqno (3.12)
$$
On using eqs. (3.2)--(3.5), it is 
then possible to show {\it by inspection} that the complete set of
equations corresponding to the secondary constraints of the
Hamiltonian formalism are (3.6)--(3.9),
and the nine equations 
$$
D_{i} \; {\widetilde p}^{\;\, ij}_{\phantom{\, ij}
\hat k\hat l}=0 \;\;\;\;i,j\ne 0 \; ,
\eqno (3.13)
$$
since these equations do not depend on time derivatives when the
$\alpha_{\hat l}$ are set to zero (see eq. (3.4)).
 
Thus, the multimomentum map does not provide all the constraints, but 
only a subset of them. To make further
progress, it is necessary to compare the
set of constraints obtained here with those found in the corresponding
Hamiltonian approach [4,6].  
\vskip 0.3cm
\leftline {\bf 4. - Self-dual gravity.}
\vskip 0.3cm
\noindent
In refs.[4,6] the Hamiltonian formulation of a complex self-dual action
on a null hypersurface in Lorentzian space-time 
was studied. The 3+1 decomposition was inserted into
the Lagrangian, and the constraints were derived with the usual
Dirac's procedure. In this section the 
results of ref.[4] are briefly
summarized and then compared with the corresponding constraints obtained by
the multimomentum map. Since the constraints found in 
Lagrangian formalism correspond to the secondary constraints of the
Hamiltonian formalism [11], the discussion is focused on these ones. 
 
The complex self-dual part of the connection are the complex 
one-forms given by
$$
{ }^{(+)}\omega_a^{\phantom{a}\hat a\hat c} =
{1\over2}\left( \omega^{\phantom{a}\hat a\hat c}_a - {i\over 2}
\epsilon^{\hat a\hat c}_{\phantom{ac} \hat b\hat d}\,
\omega_a^{\phantom{a}\hat b\hat d} \right) \; .
\eqno (4.1)
$$
Explicitly, one has
$$
{ }^{(+)}\omega^{\phantom{a}\hat 0\hat 1 }_{a}
={ }^{(+)}\omega^{\phantom{a}\hat 2\hat 
3 }_{a}={1\over2}\left( 
\omega^{\phantom{a} \hat 0\hat 1}_{a} 
+ \omega^{\phantom{a} \hat 2\hat 3}_{a}
\right) \; ,
\eqno (4.2)
$$
$$
{ }^{(+)}\omega^{\phantom{a}\hat 2\hat 1 }_{a}= 
\omega^{\phantom{a} \hat 2\hat 1}_{a} \; , \;
{ }^{(+)}\omega^{\phantom{a}\hat 0\hat 3 }_{a} = 
\omega^{\phantom{a} \hat 0\hat 3}_{a} \; , \;
{ }^{(+)}\omega^{\phantom{a} \hat 0\hat 2}_{a}=
{ }^{(+)}\omega^{\phantom{a}\hat 1\hat 3}_{a}=0 \; .
\eqno (4.3)
$$
The curvature of a self-dual connection is equal to the self-dual 
part of the curvature:
$$
\Omega({ }^{(+)}\omega)={ }^{(+)}\Omega(\omega) \; .
\eqno (4.4)
$$
Thus, the complex self-dual action to be considered is [8]
$$
S_{SD}\equiv {1\over2}\int_M d^4x\,e\,e^a_{\hat a} \,e^b_{\hat b} \,
{ }^{(+)}\Omega^{\phantom{ab}\hat a\hat b}_{ab} \; .
\eqno (4.5)
$$
Unlike the previous sections, the tetrad vectors occur in the 
following equations in a combination 
which is the self-dual part of eq. (2.10).
The 13 secondary constraints obtained in the Hamiltonian formalism in
ref.[4] after the 3+1 split, 
and written with the notation of the
present paper, are as follows: 
$$
{\cal H}_0\equiv 
-\left({e\over N}\right)^{2} V_{\hat 2}^{i}
\left[{ }^{(+)}\Omega_{ij}^{\phantom{ij}\hat 
0\hat 1}
V_{\hat 3}^{j}
+ { }^{(+)}\Omega_{ij}^{\phantom{ij}\hat 2\hat 1}V_{\hat 1}^{j}\right]
\approx 0 \; ,
\eqno (4.6)
$$
$$
{\cal H}_{i} \equiv {e\over N}\left[
{ }^{(+)}\Omega_{ij}^{\phantom{ij}\hat 0\hat 1}V_{\hat 1}^{j}+
{ }^{(+)}\Omega_{ij}^{\phantom{ij}\hat 0\hat 3}V_{\hat 3}^{j}
\right] \approx 0 \; ,
\eqno (4.7)
$$
$$
{\cal G}_{\hat 1} \equiv
-\pd_{i}\left({e\over N}V^{i}_{\hat 1}\right) 
-2{e\over N} { }^{(+)}\omega_{i}^{\; \; {\hat 0}{\hat 3}} 
V_{\hat 3}^{i} \approx 0 \; ,
\eqno (4.8)
$$
$$
{\cal G}_{\hat 2}\equiv
-{e\over N}{ }^{(+)}\omega_{i}^{\phantom{i}
\hat 0\hat 3}V^{i}_{\hat 1}\approx 0 \; , 
\eqno (4.9)
$$
$$
{\cal G}_{\hat 3}\equiv
-\pd_{i}\left({e\over N}V^{i}_{\hat 3}\right) 
+{e\over N} { }^{(+)}\omega_{i}^{\phantom{i}
\hat 2\hat 1}V_{\hat 1}^{i} 
+2{e\over N}{ }^{(+)}\omega_{i}^{\phantom{i}\hat 0\hat 1}
V^{i}_{\hat 3}\approx 0 \; ,
\eqno (4.10)
$$
$$
\eqalignno{
\chi^{i} & \equiv
-2\pd_{j}\left({e^{2}\over N} 
V_{\hat 2}^{[i}V_{\hat 1}^{j]}\right) 
- 2 {e^{2}\over N} \; 
{ }^{(+)}\omega^{\phantom{i}\hat 0\hat 3}_{j} 
V_{\hat 2}^{[i} V_{\hat 3}^{j]}
- 4 {e^{2}\over N} \; 
{ }^{(+)}\omega^{\phantom{i}\hat 0\hat 1 }_{j} 
V_{\hat 2}^{[i} V_{\hat 1}^{j]} 
+ 2 {e\over N} \; 
{ }^{(+)} \omega^{\phantom{i}\hat 0\hat 3}_{j} 
N^{[i}  V_{\hat 1}^{j]} \cr
& + {e\over N} \; 
{ }^{(+)}\omega^{\phantom{i}\hat 0\hat 3 }_{0} V_{\hat 1}^{i}
\approx 0 \; ,
&(4.11)\cr}
$$
$$
\phi_{i}\equiv -{e\over N} \left[
{ }^{(+)}\Omega^{\phantom{ij}\hat 0\hat 1}_{ij} V_{\hat 3}^{j} + 
{ }^{(+)}\Omega^{\phantom{ij}\hat 2\hat 1}_{ij} V_{\hat 1}^{j}
\right] \approx 0 \; .
\eqno (4.12)
$$
The irreducible second-class constraints turn out to be
${\cal H}_{0},{\cal G}_{\hat 3},\chi^{i},\phi_{i}V_{\hat 2}^{i}$
and $\phi_{i}V_{\hat 3}^{i}$ [4]. Note that, following refs.[4,6],
we have set to zero all the $\alpha$ parameters in the course
of deriving eqs. (4.6)--(4.12).
 
Let us now discuss the constraints from the Lagrangian point of
view. The multimomentum map is formally the same as in the Palatini case,
provided that the connection and curvature terms are replaced
by their self-dual components. Hence one has
$$ \eqalignno{
I_{{\cal S}_N}^{+}[\xi,\lambda]&=\int_{{\cal S}_N}\left[
{ }^{(+)}{\widetilde p}_{\; \; \; \; {\hat b}{\hat d}}^{\; ac}
\biggr(\xi_{\; \; ,a}^{b}
\; { }^{(+)}\omega_{b}^{\; \; {\hat b}{\hat d}}
-(D_{a} { }^{(+)}\lambda)^{{\hat b}{\hat d}}
+{ }^{(+)}\omega_{a \; \; \; \; ,b}^{\; \; {\hat b}{\hat d}}
\; \xi^{b}\biggr) \right. \cr
&\left. +{1\over 2} 
{ }^{(+)}{\widetilde p}_{\; \; \; \; {\hat b}{\hat d}}^{\; ab}
\; { }^{(+)}\Omega_{ab}^{\; \; \; {\hat b}{\hat d}}
\; \xi^{c}\right]d^{3}x_{c} \; .
&(4.13)\cr}
$$
The constraint equations obtained from setting to zero 
this multimomentum map are then
(cf. eqs. (3.6)--(3.9))
$$
\int_{{\cal S}_N} { }^{(+)}\lambda^{{\hat 0}{\hat 1}}
\left[\pd_{i}
\Bigr({e\over N}V_{\hat 1}^{i}\Bigr)
+ {e\over N}
{ }^{(+)}\omega_{i \hat 1}^{\phantom{i \hat 1}\hat 3}V_{\hat 3}^{i}
\right]d^{3}x_{0} =0 \; ,
\eqno (4.14)
$$
$$
\int_{{\cal S}_N} { }^{(+)}\lambda^{{\hat 0}{\hat 3}}
\left[\pd_{i}
\Bigr({e\over N}V_{\hat 3}^{i}\Bigr)
+ {e\over N}
{ }^{(+)}\omega_{i \hat 3}^{\phantom{i \hat 3}\hat l}V_{\hat l}^{i}
\right]d^{3}x_{0} =0 \; ,
\eqno (4.15)
$$
$$
\int_{{\cal S}_{N}} { }^{(+)}\lambda^{{\hat 1}{\hat 2}} D_{i}
{ }^{(+)}{\widetilde p}_{\; \; \; {\hat 1}{\hat 2}}^{\; 0i}
\; d^{3}x_{0}=0 \; ,
\eqno (4.16)
$$
$$
\int_{{\cal S}_{N}} \left \{ 
e V_{\hat 2}^{i}\left[ 
{ }^{(+)}\Omega^{\phantom{ij} \hat 0\hat 1}_{ij} 
V_{\hat 3}^{j}
+{ }^{(+)}\Omega^{\phantom{ij} \hat 
2\hat 1}_{ij} V_{\hat 1}^{j}\right]
-{2e\over N}N^{i} 
\left[{ }^{(+)}\Omega^{\phantom{ij} \hat 0\hat 1}_{ij}
V_{\hat 1}^{j} 
+ { }^{(+)}\Omega^{\phantom{ij} \hat 0\hat 3}_{ij} 
V_{\hat 3}^{j}
\right] \right \} \xi^{0} \, d^{3}x_{0} = 0 \; ,
\eqno (4.17)
$$
$$
\int_{{\cal S}_N} 
{e\over N} \left[
{ }^{(+)}\Omega^{\phantom{ij} \hat 0\hat 1}_{ij} 
V_{\hat 1}^{i}
+ { }^{(+)}\Omega^{\phantom{ij} \hat 0\hat 3}_{ij} 
V_{\hat 3}^{i}
\right]\xi^{j}\, d^{3}x_{0} = 0 \; .
\eqno (4.18)
$$

On the other hand,
the Euler-Lagrange equations resulting from the action (4.5) are
(cf. eqs. (3.11) and (3.12))
$$
e^{b}_{\;\hat b}\left[{ }^{(+)}\Omega^{\phantom{bh}\hat b\hat c}_{bh}-
{1\over2}e^{d}_{\;\hat a}e^{\hat c}_{\; h}
{ }^{(+)}\Omega^{\phantom{bd}\hat b\hat a}_{bd}\right]=0 \; ,
\eqno (4.19)
$$
and
$$
D_{b}{ }^{(+)}{\widetilde p}^{\;\, ab}_{\phantom{\;\,ab} 
\hat a \hat b}=0 \; .
\eqno (4.20)
$$
The self-dual Einstein equations in vacuum can be thus written 
explicitly in the form
$$
{ }^{(+)}G^{\hat 0}_{h} \equiv 
e^{b}_{\hat 1} \; 
{ }^{(+)}\Omega_{bh}^{\phantom{bh}\hat 1\hat 0} +
e^{b}_{\hat 3} \; 
{ }^{(+)}\Omega_{bh}^{\phantom{bh}\hat 3\hat 0} = 0 \; ,
\eqno (4.21)
$$
$$
{ }^{(+)}G^{\hat 1}_{h} \equiv
e^{b}_{\hat 0} \; 
{ }^{(+)}\Omega_{bh}^{\phantom{bh}\hat 0\hat 1} +
e^{b}_{\hat 2} \; 
{ }^{(+)}\Omega_{bh}^{\phantom{bh}\hat 2\hat 1} = 0 \; ,
\eqno (4.22)
$$
$$
{ }^{(+)}G^{\hat 2}_{h} \equiv
e^{b}_{\hat 1} \; 
{ }^{(+)}\Omega_{bh}^{\phantom{bh}\hat 1\hat 2} +
e^{b}_{\hat 3} \; 
{ }^{(+)}\Omega_{bh}^{\phantom{bh}\hat 3\hat 2} = 0 \; ,
\eqno (4.23)
$$
$$
{ }^{(+)}G^{\hat 3}_{h} \equiv
e^{b}_{\hat 0} \; 
{ }^{(+)}\Omega_{bh}^{\phantom{bh}\hat 0\hat 3} +
e^{b}_{\hat 2} \; 
{ }^{(+)}\Omega_{bh}^{\phantom{bh}\hat 2\hat 3} = 0 \; .
\eqno (4.24)
$$
It is easy to show that the equations independent of time
derivatives on a null hypersurface are the spatial components of eqs. 
(4.21) and (4.23), jointly with the equations
$$
D_{i}{ }^{(+)}{\widetilde p}_{\; \; \; {\hat 0}
{\hat 1}}^{\; 0i}
=D_{i}{ }^{(+)}{\widetilde p}_{\; \; \; {\hat 0}
{\hat 3}}^{\; 0i}=0 \; ,
$$
$$
D_{i}{ }^{(+)}{\widetilde p}_{\; \; \; {\hat 1}
{\hat 2}}^{\; 0i}=0 \; ,
$$
which are equivalent to (4.14)--(4.16), and (cf. eq. (3.7))
$$
D_{j}{ }^{(+)}{\widetilde p}_{\; \; \; {\hat 1}
{\hat 2}}^{\; ij}=0 \; .
\eqno (4.25)
$$
The comparison of eqs. (4.6)--(4.12) with eqs. (4.14)--(4.18) shows
that eq. (4.14) corresponds to eq. (4.8), 
eq. (4.15) to eq. (4.10), eq. (4.16) to eq. (4.9),
eq. (4.17) to eqs. (4.6) and (4.7), and eq. (4.18) 
to eq. (4.7). 

Interestingly, the constraint equations (4.15) and (4.17),
which are second class in the Hamiltonian formalism,
contribute to the constraint set of the multimomentum map,
and hence should be regarded as first class [7]. Indeed,
if one breaks covariance, so that the internal rotation
group $O(3,1)$ is replaced by 
its subgroup $O(3)$, one can set
$\lambda^{{\hat 0}{\hat k}}=0, \forall {\hat k}=1,2,3$.
After doing this, ${ }^{(+)}\lambda^{{\hat 0}{\hat 3}}$
vanishes and Eq. (4.15) reduces to an identity, while
${ }^{(+)}\lambda^{{\hat 0}{\hat 1}}$ and
${ }^{(+)}\lambda^{{\hat 1}{\hat 2}}$ remain different
from zero, so that only eqs. (4.14) and (4.16) survive.
As far as eq. (4.17) is concerned,
one should note that, 
in a 3+1 split of space-time, 
$Diff(M)$ is replaced by its subgroup
$Diff({\cal S}_{N}) \times Diff(\Re)$. In this case, barring 
some mathematical rigour, one can say that
the arbitrary vector field $\xi$ admits the decomposition
$\xi^{a}=\xi^{a}_{||}
+{\xi_{\perp}}^{a}$, where 
$\xi^{a}_{||}$ represent the infinitesimal 
three-dimensional diffeomorphisms on a generic 
hypersurface $\Sigma$ (for the time being, $\Sigma$
can be either spacelike or null),
and ${\xi_{\perp}}^{a}$ represent the 
diffeomorphisms ``off" this hypersurface. It should
be noticed that $\xi^{a}_{\perp}$ is a vector 
proportional to $n^{a}=g^{ab}f_{;b}$, where $f=0$ is the equation 
which defines locally the hypersurface. Strictly, the normal
is a field on $\Sigma$. However, in a neighbourhood of
$\Sigma$ one can introduce a slicing of $M$ viewed as the
Cartesian product $I \times \sigma(r)$, where $I$ is the
closed interval $[0,\varepsilon]$ and $\sigma$ is a three-dimensional
hypersurface obtained by moving the points of $\Sigma$
along the normal geodesics to the distance $r$, so that
$\sigma(0)$ coincides with $\Sigma$. This procedure
makes it possible to extend all tensor fields defined on
$\Sigma$, including the normal, to tensor fields on $M$.

In adapted coordinates, one has $\xi_{||}^{0}=0$ and
$n^{a}=g^{a0}\partial_{0} f$. By virtue of (2.9)
it follows that, 
on a null hypersurface, $\xi_{\perp}^{0}$ vanishes as well.
This in turn implies that eq. (4.17) 
reduces to an identity.
 
These simple remarks seem to point out to a deeper 
interpretation of our result: eqs. (4.15) and (4.17) may or may not
be considered first-class constraints, depending on 
whether or not one breaks the original diffeomorphism 
group of the theory to a proper subgroup. In other words,
only when $Diff(M)$ and $O(3,1)$ are replaced by their
subgroups $Diff({\cal S}_{N}) \times Diff(\Re)$ and
$O(3)$, the constraints (4.15) and (4.17) become second
class and hence do not contribute to the multimomentum map.
\vskip 0.3cm
\leftline {\bf 5. - Concluding remarks.}
\vskip 0.3cm
\noindent
This paper has considered the application of the multimomentum-map
technique to study general relativity as a constrained
system on null hypersurfaces. Its contribution lies in relating
different formalisms for such a constraint analysis.
We have found that, on a null hypersurface, the multimomentum map
provides just a subset of the full set of constraints, regarded as those
particular Euler-Lagrange equations which are not of evolutionary type. 

Although the multimomentum map is expected to yield only
the secondary first-class constraints [8], we have found that
some of the constraints which are second class in the
Hamiltonian formalism occur also 
in the Lagrangian multimomentum
map. This leads to inequivalent formalisms. 
Such inequivalence can be interpreted observing that
our analysis remains covariant in that it
deals with the full diffeomorphism group of space-time,
say $Diff(M)$, jointly with the internal rotation group
$O(3,1)$. Hence one incorporates some constraints which are
instead ruled out if one breaks covariance, which amounts
to taking subgroups of the ones just mentioned (see sect.
{\bf 4}). The remaining (second-class) 
constraints have been found just by 
checking which Euler-Lagrange equations 
are not of evolutionary type.
This property suggests, perhaps, that second-class constraints can 
be treated by introducing some modifications in the construction
of the multimomentum map. The work in ref.[7] shows
that a suitable definition of ``momentum map" may be introduced,
so as to incorporate the analysis of primary first-class
constraints as well. Of course, this is no longer a Lagrangian
analysis [11], but appears to be an important issue for
further research.

Another relevant problem lies in the
constraint analysis on double null hypersurfaces. These consist
of two null hypersurfaces, intersecting each other
in a spacelike two-surface [5,12]. On a double null hypersurface,
the integrations $d^{3}x_{0}$ and $d^{3}x_{1}$ (cf. eq. (3.1))
both survive in the multimomentum-map equations.

The multisymplectic framework 
appears to have very interesting 
features both in general relativity and in other field theories [7],  
but our paper shows that there is still an unsatisfactory state of 
affairs in this formalism because of the lack of a systematic algorithm to 
generate all constraints of the theory, 
since they have been found just by 
inspection of the field equations. 
Perhaps one needs a suitable version of covariant
Hamiltonian formalism for constrained systems (cf. refs.[13,14]).
A proper understanding of all the above issues will show,
presumably, whether or not the multimomentum-map 
formalism offers substantial advantages with respect 
to the well established Hamiltonian techniques [1-4,6].
\vskip 0.3cm
\centerline {$* * *$}
\vskip 0.3cm
We are indebted to Mark Gotay and Luca Lusanna 
for useful discussions, and to Giuseppe Marmo
for encouraging our work on null hypersurfaces. It is also
a pleasure to thank Gabriele Gionti, with whom we shared
some years of work on constrained systems. Last,
but not least, we are grateful to Joshua Goldberg, 
whose ideas on canonical gravity inspired our research.
\vskip 0.3cm
\leftline {REFERENCES}
\vskip 0.3cm
\item {[1]}
GOLDBERG J. N., {\it Found. Phys.}, {\bf 14} 
(1984) 1211.
\item {[2]}
NAGARAJAN R. and GOLDBERG J. N., {\it Phys. Rev. D},
{\bf 31} (1985) 1354.
\item {[3]}
GOLDBERG J. N., {\it D-invariance on a null surface}, in
{\it Gravitation and Geometry}, edited by W. RINDLER and
A. TRAUTMAN (Bibliopolis, Naples) 1987, p. 159.
\item {[4]}
GOLDBERG J. N., ROBINSON D. C. and SOTERIOU C., 
{\it Class. Quantum Grav.}, {\bf 9} (1992) 1309.
\item {[5]}
d'INVERNO R. A. and VICKERS J. A., {\it Class. Quantum Grav.},
{\bf 12} (1995) 753.
\item {[6]}
GOLDBERG J. N. and SOTERIOU C., {\it Class. Quantum Grav.}, 
{\bf 12} (1995) 2779.
\item {[7]}
GOTAY M. J., ISENBERG J., MARSDEN J. E. and MONTGOMERY R.,
{\it Momentum Mappings and the Hamiltonian Structure of
Classical Field Theories with Constraints} 
(Springer, Berlin) in press.
\item {[8]}
ESPOSITO G., GIONTI G. and STORNAIOLO C., 
{\it Nuovo Cimento B}, {\bf 110} (1995) 1137.
\item {[9]}
ESPOSITO G. and STORNAIOLO C., {\it Class. Quantum Grav.},
{\bf 12} (1995) 1733.
\item {[10]}
ESPOSITO G. and STORNAIOLO C., {\it Nuovo Cimento B},
{\bf 111} (1996) 271.
\item {[11]}
ESPOSITO G., GIONTI G., MARMO G. and STORNAIOLO C.,
{\it Nuovo Cimento B}, {\bf 109} (1994) 1259.
\item {[12]}
SACHS R. K., {\it J. Math. Phys.}, {\bf 3}
(1962) 908.
\item {[13]}
D'ADDA A., NELSON J. E. and REGGE T., 
{\it Ann. Phys. (N.Y.)}, {\bf 165} (1985) 384.
\item {[14]}
NELSON J. E. and REGGE T., {\it Ann. Phys. (N.Y.)},
{\bf 166} (1986) 234.

\bye